\documentclass[3p,times,twocolumn]{elsarticle}
 \biboptions{comma,sort&compress}
 
\usepackage{graphicx}
\usepackage{here}
\usepackage{ecrc}

\volume{00}

\firstpage{1}

\journalname{Nuclear and Particle Physics Proceedings}

\runauth{}

\jid{nppp}

\jnltitlelogo{Nuclear and Particle Physics Proceedings}

\usepackage{amssymb}

\usepackage[figuresright]{rotating}
\usepackage{hyperref}

\begin{document}

\begin{frontmatter}

\title{ Measurements of heavy-flavor jets with ALICE at the LHC
 $^*$}
 \cortext[cor0]{Talk given at 22nd High Energy Physics International Conference in Quantum Chromodynamics (QCD 19), 2 - 5 July 2019, Montpellier - FR}
 \author{Ashik Ikbal Sheikh (for the ALICE Collaboration)}

\ead{ashikhep@gmail.com}
\address{Variable Energy Cyclotron Centre, Kolkata - 700064, India}
\address{Homi Bhabha National Institute, Mumbai - 400094, India}

\pagestyle{myheadings}
\markright{ }
\begin{abstract}
Heavy quarks created in ultra-relativistic heavy-ion collisions are mostly produced in hard QCD processes during the early stages of the reaction. They interact with the hot and cold nuclear matter throughout the evolution of the medium via semi-hard and soft processes such as energy loss via gluon radiations and collisions. Nuclear modification of heavy flavors in p-A systems provides insight into cold nuclear matter effects such as (anti)shadowing and $k_{T}$-broadening, and serves as a baseline for A-A studies. In addition to that the fully reconstructed heavy-flavor jets provide additional information on the flavor (or mass) dependence of fragmentation, color charge effects as well as insight into the contribution of late gluon splitting. In this contribution, we present the measurements of $b$-jet production in p--Pb collisions at $\sqrt{s_{\rm NN}}=5.02$ TeV and $c$-jet production in pp, p--Pb and Pb--Pb collisions measured by the ALICE experiment at the LHC. The measurements of the nuclear modification factors for $c$-jet in p--Pb and Pb--Pb collisions are also presented. The experimental measurements are compared with the results from Monte Carlo event generators (PYTHIA 6, PYTHIA 8 and Herwig 7) and the NLO pQCD calculations (POWHEG+PYTHIA6). We find good agreement of the measurements with the results from Monte Carlo event generators and from NLO pQCD calculations.
\end{abstract}

\end{frontmatter}
\section{Introduction}
The relativistic heavy-ion collision programs at the Relativistic Heavy Ion Collider (RHIC) at BNL and the Large Hadron Collider (LHC) at CERN aim to produce a hot and dense deconfined
state of QCD matter, called quark-gluon plasma (QGP). Many experimental results indicate that
this new deconfined state of matter has been formed during relativistic heavy-ion collisions
at the RHIC~\cite{RHIC} and the LHC~\cite{LHC}. One of the features of this deconfined plasma is the energy loss of hard partons leading to reduced yield of open-charm and open beauty mesons in Pb-Pb collisions as compared to the yield in pp collisions scaled with the number of collisions, a phenomenon known as jet quenching. 
The proton-nucleus (p-A) collisions are essential to understand the effects that take place in the cold nuclear matter (CNM),
which serve as a baseline for the measurements done in A-A collisions. The influence of the CNM effects can be studied by measuring the nuclear modification factor in p-A collisions as: 

\begin{eqnarray}
 R_{pA} = \frac{1}{A}\frac{\rm d\sigma_{pA}/dp_T}{\rm d\sigma_{pp}/dp_T} 
\end{eqnarray}
where, $\rm d\sigma_{pA}/dp_T$ and $\rm d\sigma_{pp}/dp_T$ are the $p_T$-differential  production cross sections of a given  particle species in p-A and pp collisions, respectively, and A is the number of nucleons in the nucleus. 

The usual measurements of heavy flavor mesons ($\rm D$ and $\rm B$) reconstructed either directly or via their decay electrons (HFe) provide important information about the jet quenching and collective motion of the $c$ and $b$ quarks within the medium. Reconstruction of jets containing a heavy quark also provides the information about the flavor dependence of the fragmentation mechanism. The partonic energy loss is expected to be mass dependent and the corresponding effects should be more for partons with low transverse momentum ($p_T$). 

Heavy quarks (charm ($c$) and bottom ($b$)) are mostly produced in primordial stage of the heavy-ion collisions from the initial fusion of partons. Like light quarks or gluons, heavy quarks fragment into jets, called heavy-flavor jets ($c$-jet and $b$-jet). The loss of energy in the dense medium due to heavy quarks is different from those due to light quarks and gluons, particularly in the low and intermediate transverse momentum region. Therefore, the jet quenching depends on the flavor of the fragmenting parton as discussed in Refs.~\cite{Solana,Gyulassy,Magdelena}.

Experimentally, the heavy quark content of a jet can be identified by looking for the presence of heavy-flavor hadrons inside the jet. The hadrons containing heavy quarks have sufficient lifetimes ($\sim 10^{-12}$ s), so they travel some distances ($\sim$ few $mm$) before decaying. The properties of their decay vertices allow us to identify heavy-flavor tagged jets. The CMS~\cite{cmsb,cmsc} and ATLAS~\cite{atlasb} collaborations at the LHC have measured the heavy-flavor jet production and suppression in heavy-ion collisions.

In this paper, we show the first ALICE measurements of $p_T$-differential production cross section for $b$-jets in p--Pb collisions at $\sqrt {s_{\rm NN}} = 5.02$ TeV, and for $\rm D^0$-meson tagged jets in pp collisions at $\sqrt s = 5.02, 7$ and $13$ TeV and HFe tagged jets in pp collisions at $\sqrt {s} = 5.02$ TeV. The measurements of fractional jet momentum carried by $\rm D$-meson along the jet
axis direction ($z_{||}^{ch}$) are shown for $\rm D^0$-meson tagged jet. The $z_{||}^{ch}$ is defined as:

\begin{equation}
z_{||}^{ch} = \frac{\vec{p}_{jet}.\vec{p}_{D}}{\vec{p}_{jet}.\vec{p}_{jet}}
\end{equation} 
where, $\vec{p}_{jet}$ and $\vec{p}_{D}$ are the jet momentum and $\rm D$-meson momentum respectively.
We also report the measurements of nuclear modification factor ($R_{pPb}$,$R_{AA}$) for $\rm D$-meson and HFe tagged jet in p--Pb and Pb--Pb collisions at $\sqrt {s_{\rm NN}} = 5.02$ TeV. For further details on measurements of $\rm D^0$-meson tagged jets, we refer to~\cite{AliceD0jet}. The experimental measurements are compared with results from Monte Carlo event generators (PYTHIA 6, PYTHIA 8 and Herwig 7) and the NLO pQCD calculations (POWHEG+PYTHIA6). The ALICE heavy flavor jet measurements are performed down to very low $p_T$ region, particularly for $\rm D$-meson and HFe tagged jets the measured $p_T$ range is $5<p_{T,jet}^{ch}<30$ GeV/$c$ and for $b$-jet the range is $10<p_{T,jet}^{ch}<100$ GeV/$c$. The excellent particle tracking capabilities of the ALICE detector make the low momentum heavy flavor jet measurement possible as discussed in Sec.~\ref{ALICEdet}.

The paper is organized as follows: In the next section we outline very briefly the ALICE detector setup and data sample used in the measurements. In Sec.~\ref{ana}, we describe the analysis procedure for heavy flavor jet measurements. We discuss the experimental results in Sec.~\ref{results}. Section~\ref{sumcon} contains the summary and conclusions.

\section{ALICE Detector and Data Sample}
\label{ALICEdet}
The ALICE detector~\cite{AliceDet1,AliceDet2} setup has excellent particle identification, low-$p_T$ track reconstruction and vertexing capabilities. The ALICE detector system is immersed in a longitudinal magnetic field $B=0.5$ T, produced by solenoid magnet, which bends particle trajectories to measure their momentum. The main ALICE tracking system consists of an Inner Tracking System (ITS) of 6 silicon layers used for primary and secondary vertex determination, followed by a large volume gaseous detector, Time Projection Chamber (TPC), surrounding the ITS. Beside tracking, the ITS and TPC have excellent particle identification (PID) capabilities. The particles are identified by means of specific energy loss in the TPC and time-of-flight in the TOF detector. After TPC, two full-azimuth PID detectors are present: one is Transition Radiation Detector (TRD) which is used for electron identification and another one is Time-Of-Flight (TOF) detector, used for identification of pion, kaon and proton. The large acceptance calorimeter is the Electromagnetic Calorimeter (EMCal), used
also for jet reconstruction and identification of electrons, photons and neutral pions ($\pi^0$). The identification of electron at high $p_T$ is performed with the EMCal, while TPC and TOF are used in the low $p_T$ region. 

The measurements presented here are carried out using data recorded by the ALICE detector setup. For the Monte Carlo (MC) simulations, we have used the PYTHIA6~\cite{Pythia}. The generated particles are transported through the ALICE apparatus using the GEANT3 transport model~\cite{Geant3}.

\section{Analysis Procedure}
\label{ana}
The identifications of heavy flavor jets rely on the idea of finding heavy flavor content within the jets. The heavy flavor hadrons within the jets decay after traveling some distances from the primary vertex. The identification procedure uses the information on displacement of the decay vertex from the interaction vertex.

Firstly, the jets are reconstructed from the selected charged tracks using the infrared and collinear safe anti-$k_T$ algorithm from the FastJet package~\cite{Fastjet}. The tracks are required to have $|\eta| < 0.8$, $p_T > 0.15$ GeV/c, at least 70 associated TPC space points (out of a maximum of 159), $\chi^2/ndf < 4$ in the TPC (where ndf is the number of degrees of freedom involved in the tracking procedure), at least one hit in either of the two layers of the SPD and a minimum of 3 hits in the entire ITS.

\subsection{\bf {b}-Jet Identification}
\label{bjetId}

The $b$-jet candidate is identified by means of reconstruction of a displaced secondary
vertex (SV) within the jet. The SV is reconstructed from jet constituents and is required to have 3 tracks. When there are many candidates for SV, we choose the one which is maximally displaced. Discriminating variables exploit properties of beauty-hadron decays determined by their long lifetimes
and large masses. The main discriminating variables are: (i) $SL_{xy} = L_{xy}/\sigma_{L_{xy}} >  cut\,\,\,on\,\,\,significance$, where $L_{xy}$ is the projection of the distance of the reconstructed
SV from the primary vertex on the (x, y) plane and $\sigma_{L_{xy}}$ is the resolution of $L_{xy}$. (ii) $\sigma_{\rm SV}< cut\,\,\,on\,\,\,SV\,\,\,resolution$, where the SV resolution is calculated as,
$$
\sigma_{\rm SV} = \sqrt{ \sum_{i=1}^{3}\rm d_{i}^{2} },
$$
where $\rm d_{i}$ are the closest approaches of the tracks (used to reconstruct SV) to the $\rm SV$ in 3D. The resolution, $\sigma_{\rm SV}$ has a discrimination power only if a cut on $SL_{xy}$ is applied. 

\begin{figure}[!h]
	\centering
	\begin{minipage}[b]{0.4\textwidth}
		\includegraphics[width=\textwidth]{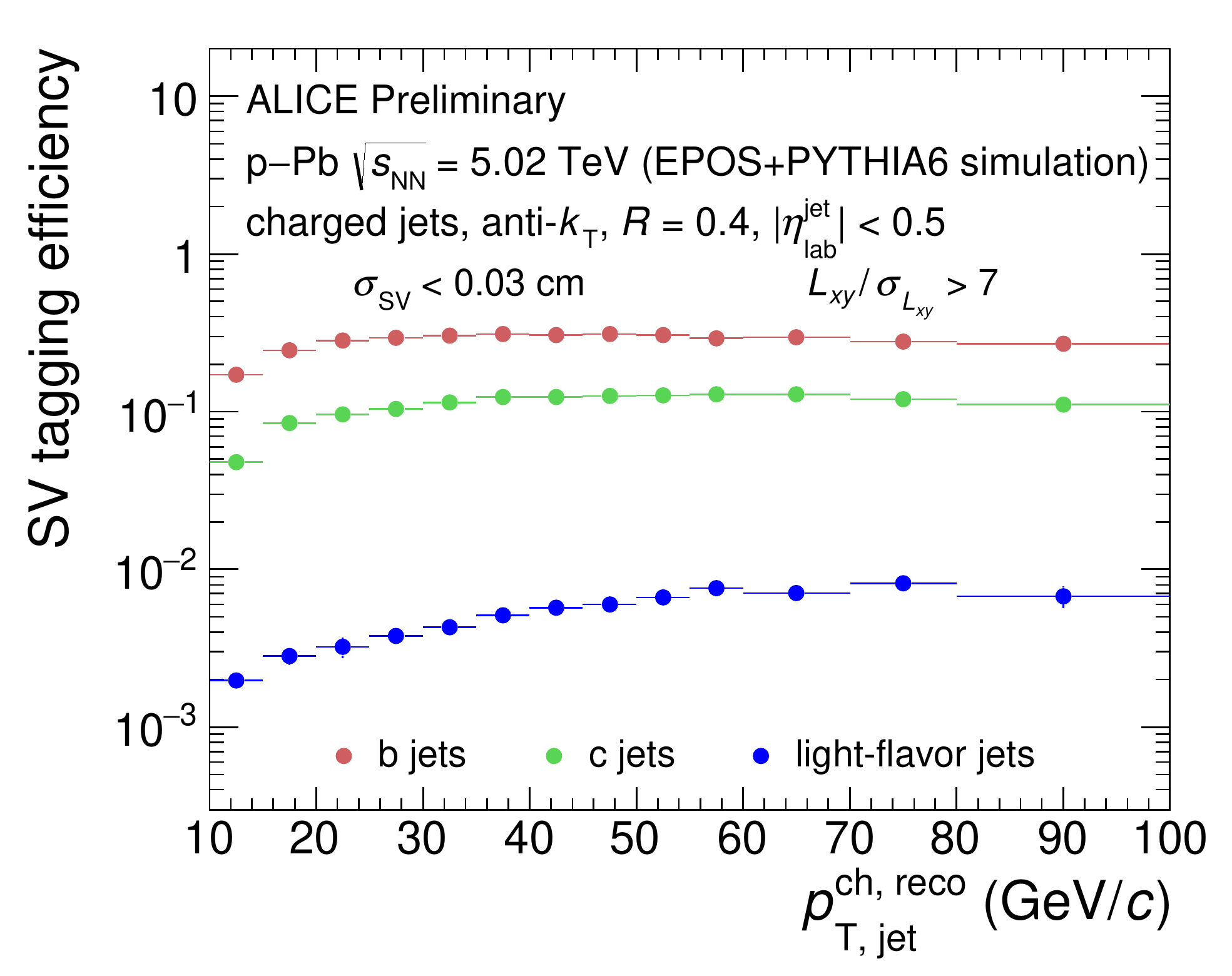}
		\caption{The $b$-jet, $c$-jet and light flavor (lf)-jet tagging efficiency as a function of jet $p_T$ for $SL_{xy} > 7$ and  $\sigma_{SV} < 0.03$ cm.}
		\label{bJetEff}
	\end{minipage}
\end{figure}

In Fig.~\ref{bJetEff}, we display the jet flavor tagging efficiency (obtained from EPOS+PYTHIA6 simulation) as a function of jet $p_T$ for $SL_{xy} > 7$ and  $\sigma_{SV} < 0.03$ cm. The tagging efficiency is defined as the number of true flavor jets ($b$-jet, $c$-jet and light flavor jet) after tagging w.r.t total number jets ($b$-jet, $c$-jet and light flavor jet) before tagging.

The primarily identified all $b$-jet candidates (using discriminating variables $SL_{xy}$ and $\sigma_{SV}$) $N_b^{all}$, are corrected for the tagging efficiency ($\epsilon_b$) and purity ($P_b$) as: $N_b = N_b^{all}P_b/\epsilon_b$. The efficiency and purity corrected $b$-jet samples are further corrected for the detector effects (unfolding) and systematic uncertainties.

\subsection{\bf {c}-Jet Identification}
\label{cjetId}

The main concept of $c$-jet tagging is the $\rm D^0$-mesons identification within the jets. The $\rm D^0$-mesons are reconstructed via their hadronic decay channel, $\rm D^0 \rightarrow K^{-} \pi^{+}$ (branching ratio $\sim$ 3.89 $\pm$ 0.04\%) and it's charge conjugate. In each event, $\rm D^0$-meson candidates and their decay vertices are constructed from pairs of tracks with opposite charge. The reconstructed $\rm D^0$-meson candidates are corrected for the reconstruction efficiency. The $\rm D^0$-mesons selection criteria have been well-established by the ALICE Collaboration as discussed in Ref.~\cite{D01,D02}. The b-hadron feed-down corrections are further done to the reconstructed $\rm D^0$-meson candidates within the jets. Finally, the jets containing the $\rm D^0$-meson is corrected for the detector effects (unfolding) and systematic uncertainties to extract the $\rm D^0$-meson tagged jet yields. More details can be seen in~\cite{Djet}.

\subsection{\bf HFe Tagged Jet Identification}
\label{cjetId}
 The specific energy loss dE/dx in the TPC volume is used for electron identification over the momentum range $0.5 < p_T < 12$ GeV/$c$. However, the electron dE/dx band intersects with the hadron band below 2.5 GeV/$c$ and merges with the hadron band above 6 GeV/$c$. The TOF and EMCal are used in the momentum range $0.5 < p_T < 2.5$ GeV/$c$ and $6 < p_T < 12$ GeV/$c$ respectively to resolve this issue. Once the electrons are identified, the next step is to subtract the non-HFe contribution where the main sources are the photonic conversion and Dalitz decay of neutral mesons. Finally, the HFe yields are obtained after performing the electron reconstruction, selection efficiency correction and correction for detector geometry. The details of the HFe identification procedure can be found in~\cite{Hfe}.

\section{Results and Discussions}
\label{results}

\begin{figure}[!h]
	\centering
	\begin{minipage}[b]{0.4\textwidth}
		\includegraphics[width=\textwidth]{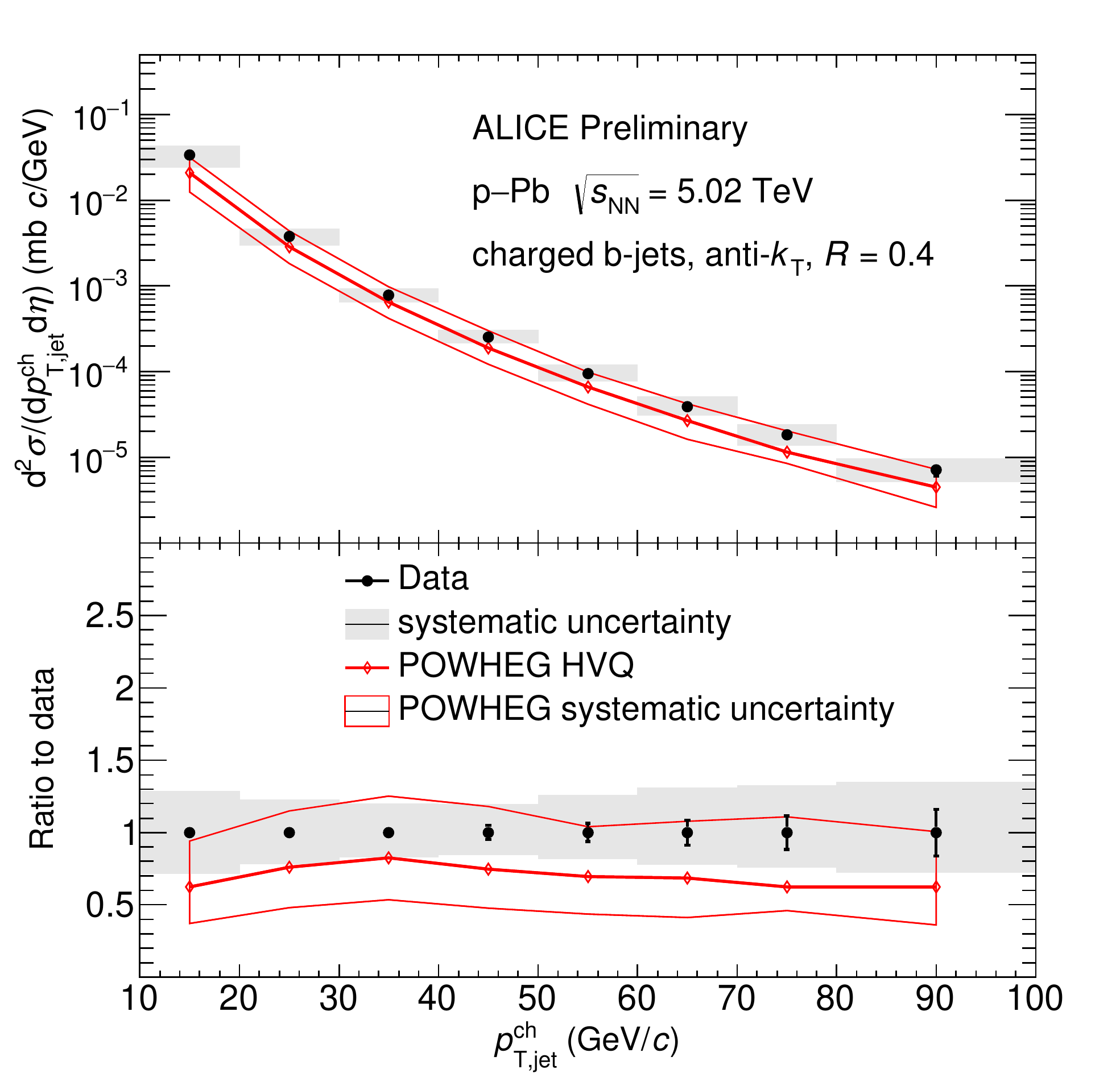}
		\caption{Upper panel: The measured $b$-jet corss-section as a function of charged jet $p_{T}$. Lower panel: The ratio of the measured $b$-jet spectra with the NLO pQCD calculations (POWHEG).}
		\label{bxsec}
	\end{minipage}
\end{figure}

The $p_T$-differential production cross section for $b$-jets with resolution parameter $R = 0.4$, reconstructed from charged particles in minimum bias p--Pb collisions at $\sqrt {s_{\rm NN}} = 5.02$ TeV is shown in Fig.~\ref{bxsec}. We have compared the measured $b$-jet cross section with NLO pQCD calculations (POWHEG+PYTHIA). The measured $b$-jet cross-section is in agreement with the NLO pQCD calculations within the experimental and theoretical uncertainties as seen in the ratio (data over calculation) plot in the lower panel of Fig.~\ref{bxsec}. 

\begin{figure}[!h]
	\centering
	\begin{minipage}[b]{0.38\textwidth}
		\includegraphics[width=\textwidth]{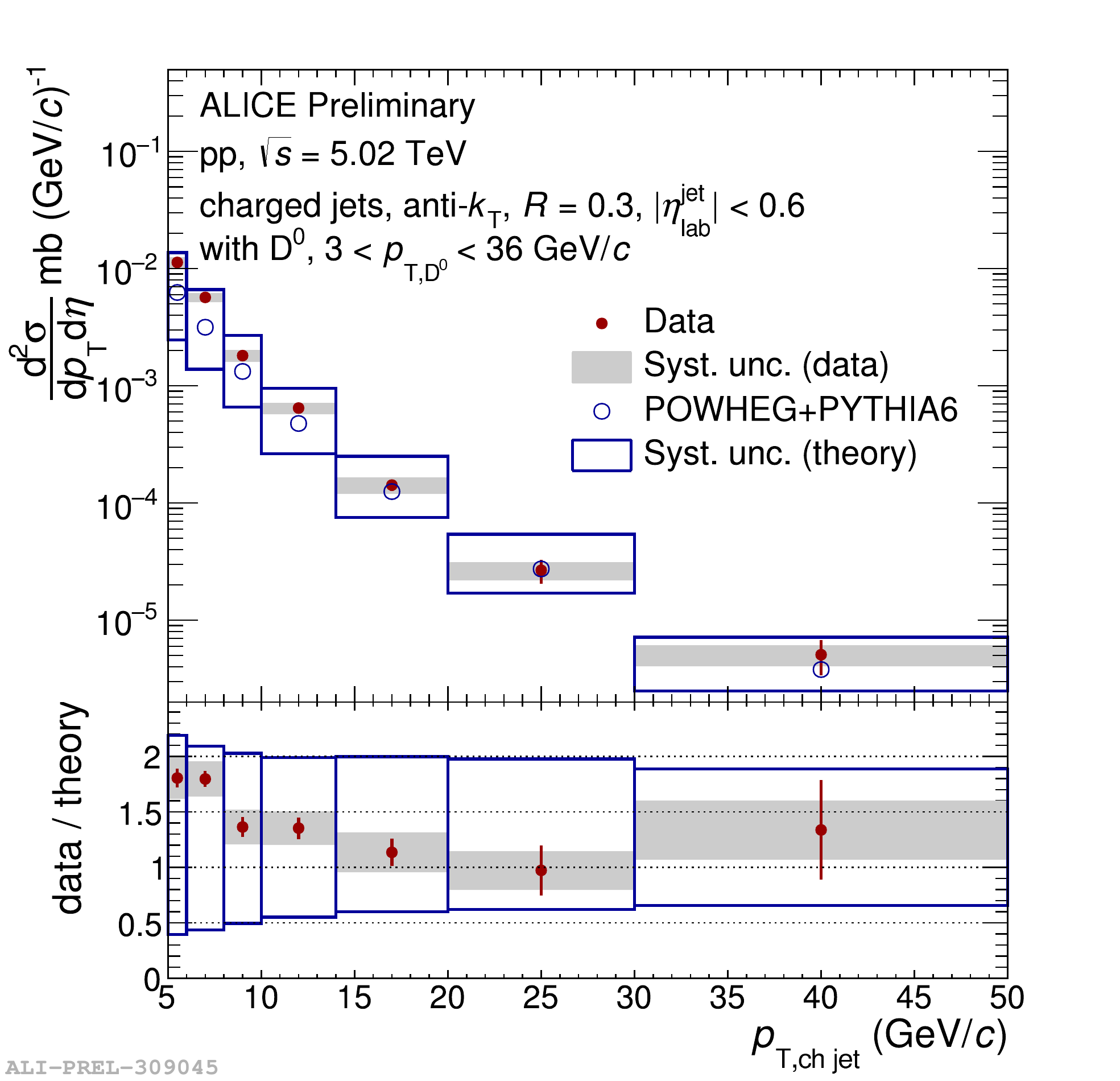}
		\caption{Charm jet (tagged with $\rm D^0$-meson) $p_T$-differential cross section in pp collisions at $\sqrt s = 5.02$ TeV, compared with POWHEG+PYTHIA6 NLO pQCD calculations.}
		\label{Djetpp5TeV}
	\end{minipage}
\end{figure}

\begin{figure}[!h]
		\centering
	\begin{minipage}[b]{0.36\textwidth}
		\includegraphics[width=\textwidth]{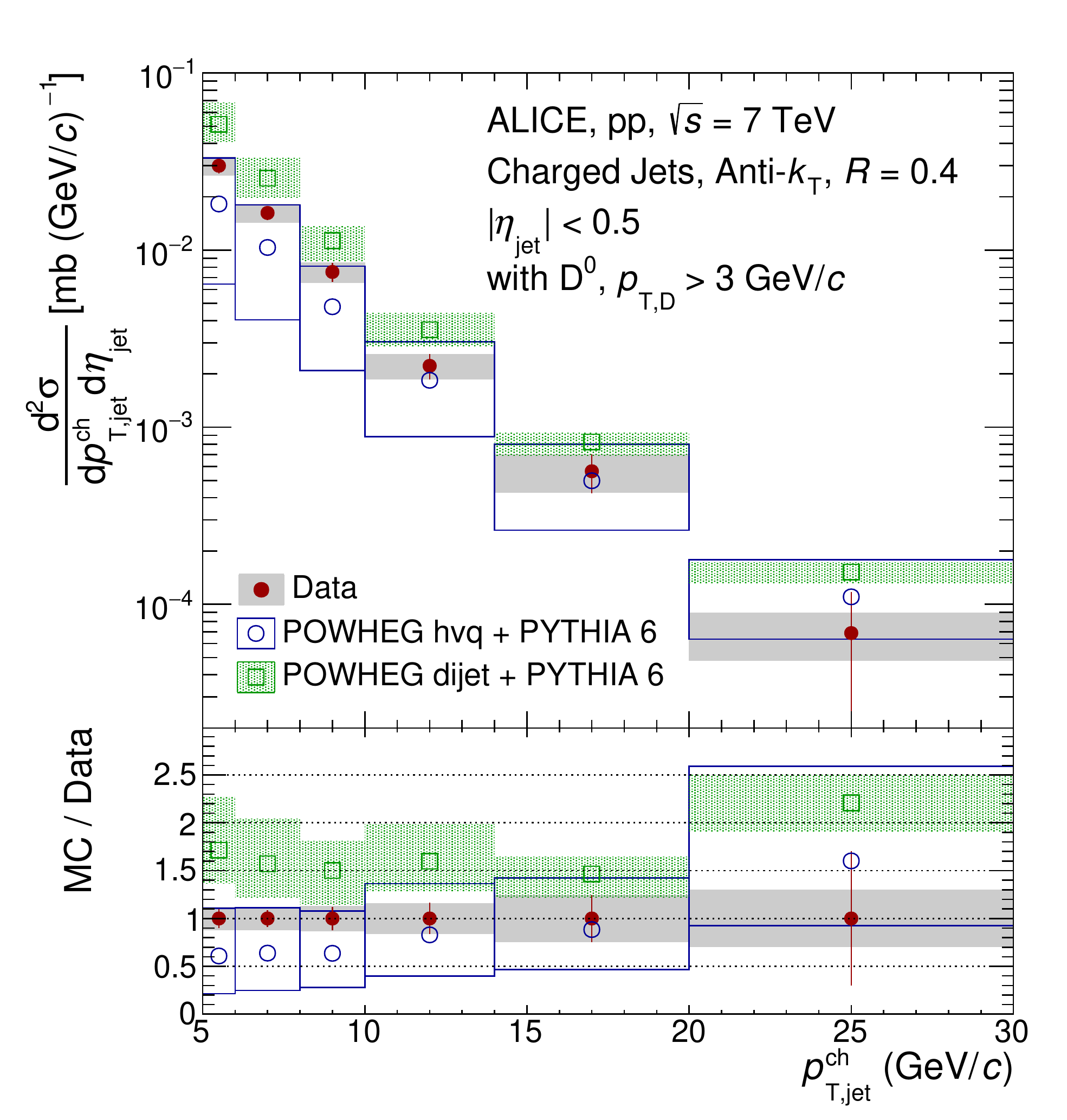}
		\caption{Same as Fig.~\ref{Djetpp5TeV} but in pp collisions at $\sqrt s = 7$ TeV.}
		\label{Djetpp7TeV}
		\includegraphics[width=\textwidth]{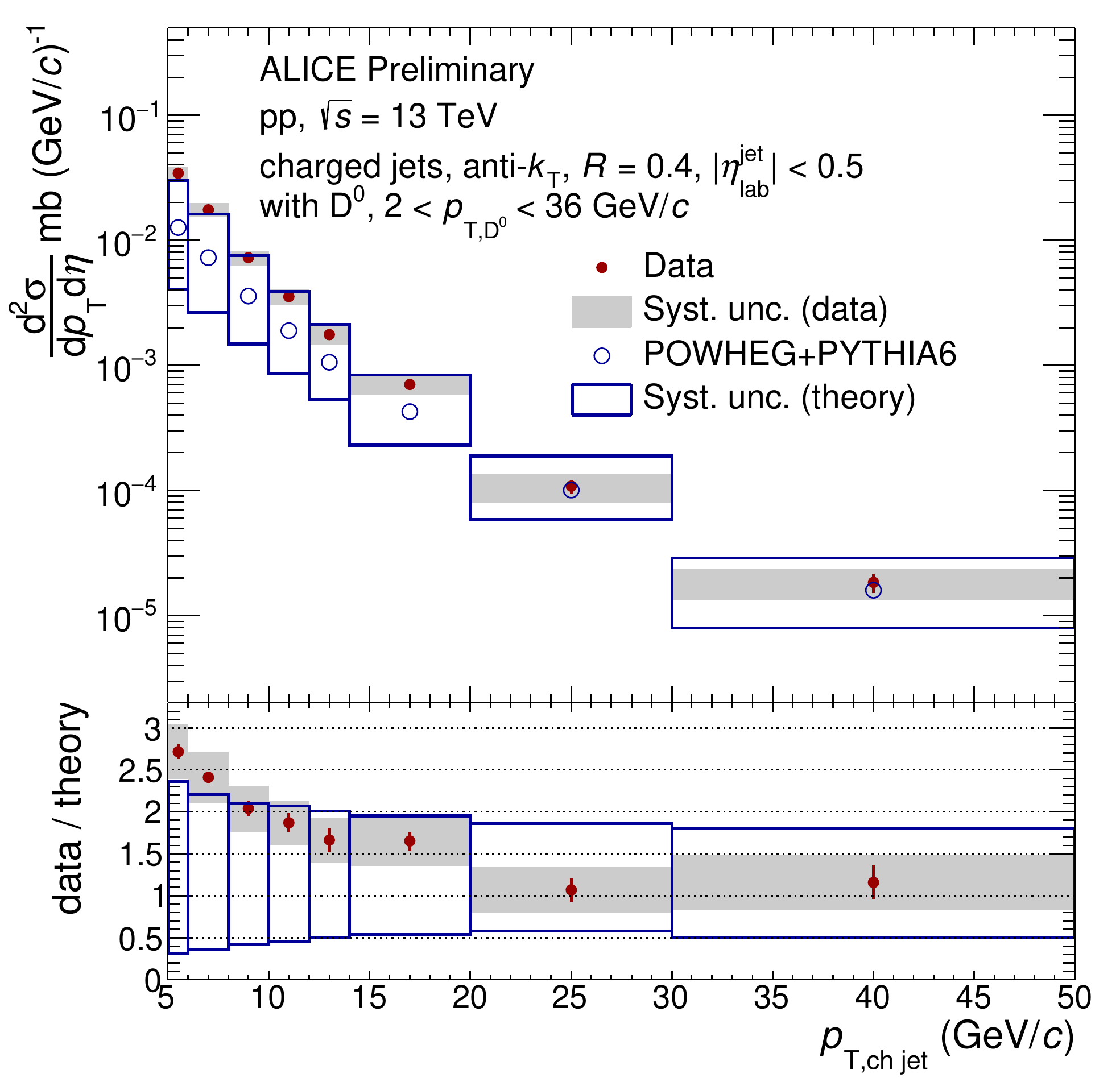}
		\caption{Same as Fig.~\ref{Djetpp5TeV} but in pp collisions at $\sqrt s = 13$ TeV.}
		\label{Djetpp13TeV}
	\end{minipage}
\end{figure}

Figures.~\ref{Djetpp5TeV}, \ref{Djetpp7TeV} and \ref{Djetpp13TeV} show the $p_T$-differential cross section of charm jets containing a $\rm D^0$-meson in pp collisions at $\sqrt s = 5.02, 7$ and $13$ TeV respectively. The $\rm D^0$-mesons used to tag the jets have a minimum transverse momentum 3 GeV/$c$ for $\sqrt s = 5.02, 7$ TeV and 2 GeV/$c$ for $\sqrt s = 13$ TeV. The measurements are compared with NLO pQCD calculations obtained with the POWHEG-BOX V2 framework~\cite{Pow1,Pow2,PowBox}, matched with PYTHIA~6 (Perugia-2011 tune)
for the generation of the parton shower and of the non-perturbative aspects of the simulation, such as hadronization of colored partons and generation of the underlying event. The theoretical uncertainties are estimated by varying the renormalization and factorization  scales ($0.5\mu_0\leq\mu_{\rm F, R}\leq2.0\mu_0$ with $0.5\leq\mu_{\rm R}/\mu_{\rm F}\leq2.0$), 
the mass of the charm quark ($m_{\rm c}=1.3,~1.7$~GeV/$c^2$\ with $m_{\rm c, 0}=1.5$~GeV/$c^2$) 
and the parton distribution function (central points: CT10nlo; variation: MSTW2008nlo68cl~\cite{Martin:2009}).
Two process implementations of the POWHEG framework are employed: the heavy-quark~\cite{Frixione:2007b} and the di-jet implementation~\cite{Alioli:2010b}.
A good agreement is found within the theoretical and 
experimental uncertainties between the measured $p_T$-differential cross section and the cross section obtained with the POWHEG heavy-quark implementation as shown in Figs.~\ref{Djetpp5TeV}, \ref{Djetpp7TeV} and \ref{Djetpp13TeV}. However, the POWHEG di-jet implementation
systematically overestimates the production yield at $\sqrt s = 7$ TeV by a constant factor of $\approx 1.5$.

\begin{figure}[!h]
	\centering
	\begin{minipage}[b]{0.36\textwidth}
		\includegraphics[width=\textwidth]{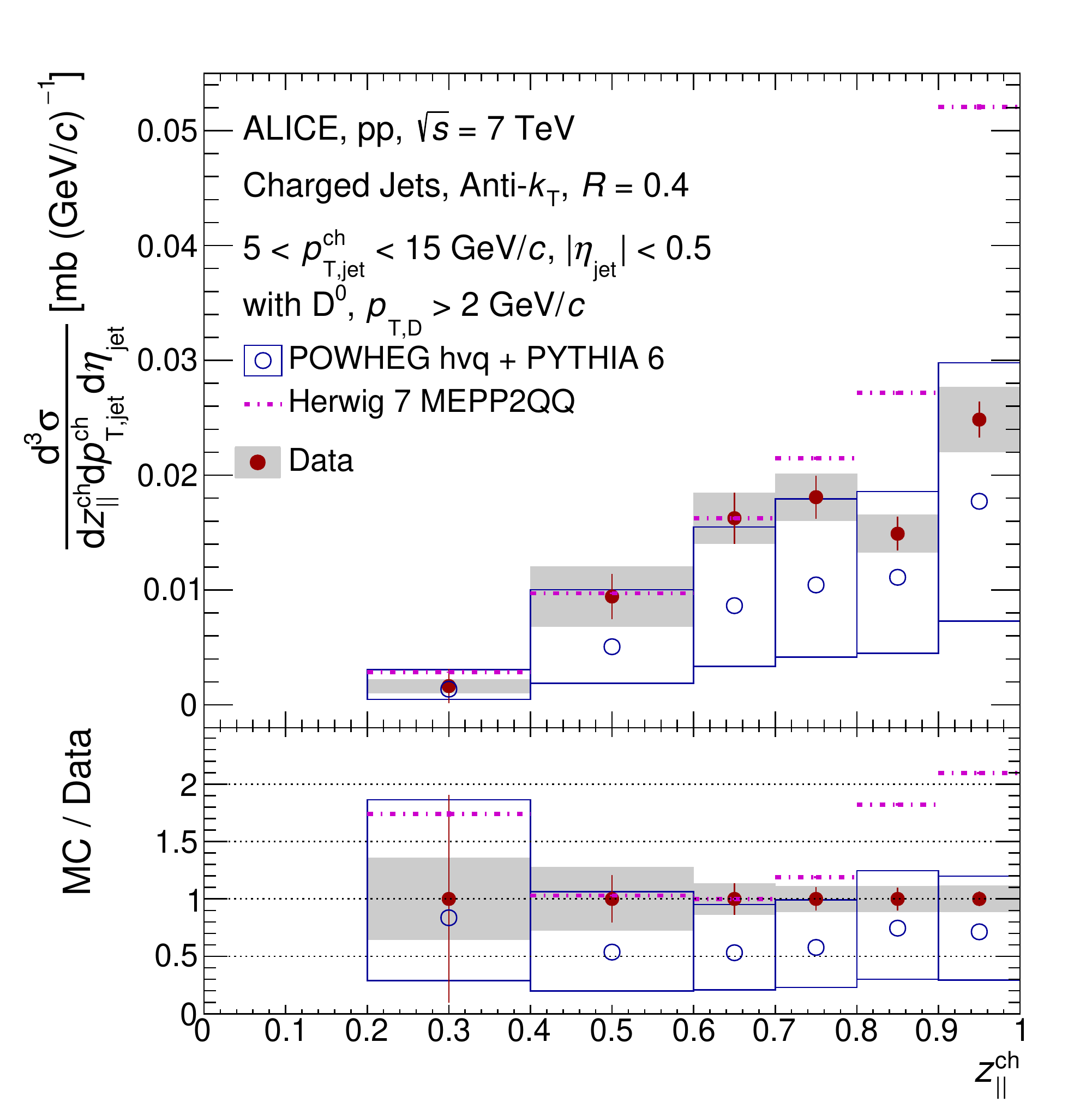}
		\includegraphics[width=\textwidth]{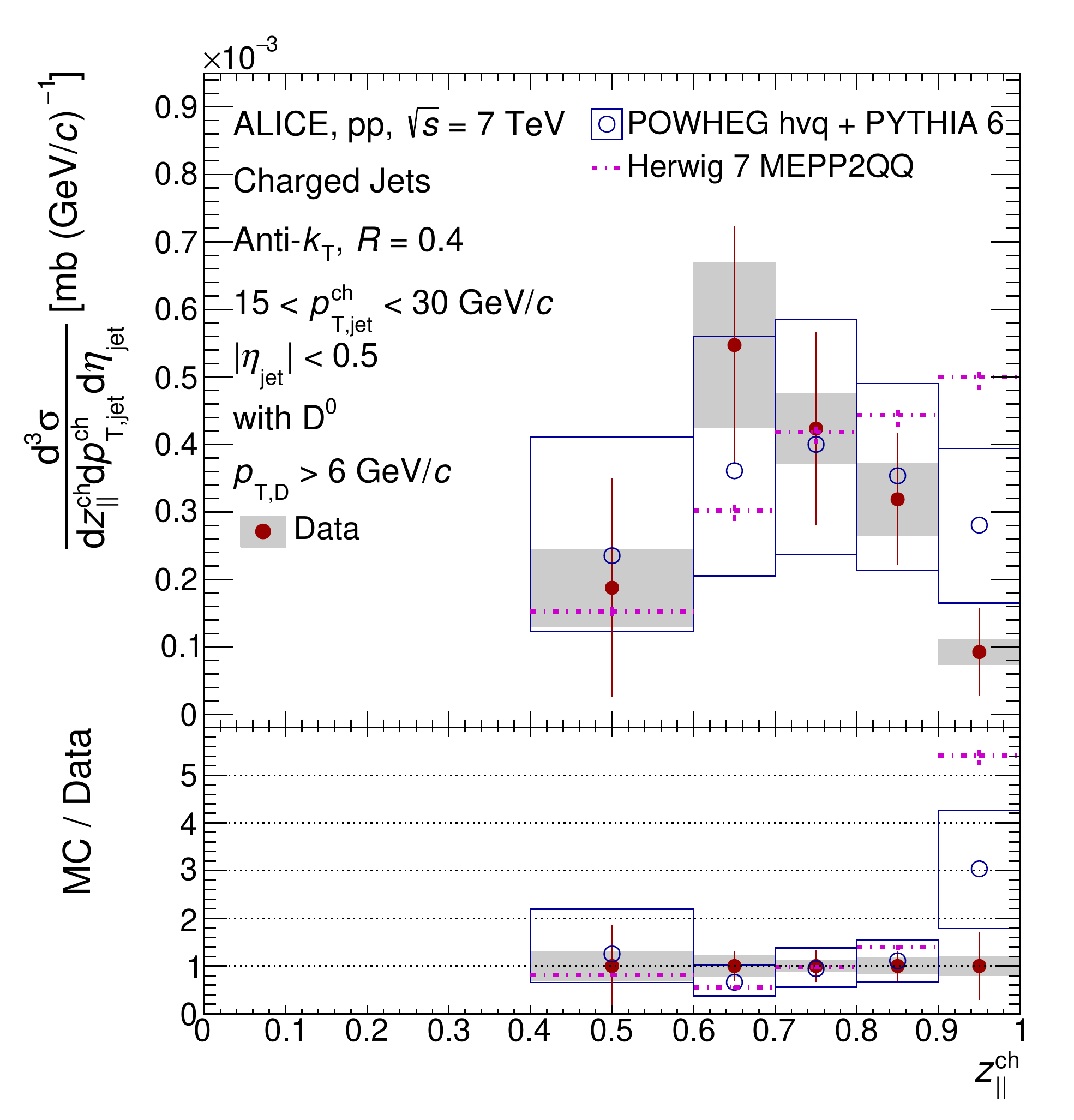}
		\caption{ The $z_{||}^{ch}$-differential cross section of Charm jet (tagged with $\rm D^0$-meson) in pp collisions at $\sqrt s = 7$ TeV with $5<p_{T,jet}^{ch}<15$ GeV/$c$ (upper plot) and $15<p_{T,jet}^{ch}<30$ GeV/$c$ (lower plot). The measurements are compared with the results of POWHEG+PYTHIA6 and Herwig 7 MEPP2QQ.}
		\label{zpDjet}
	\end{minipage}
\end{figure}

 In Fig~\ref{zpDjet}, we show the $z_{||}^{ch}$-differential cross section of $\rm D^0$-meson tagged jets for $5<p_{T,jet}^{ch}<15$~GeV/$c$ (upper plot) and for $15<p_{T,jet}^{ch}<30$~GeV/$c$ (lower plot). The measurements are compared with simulations obtained with the POWHEG heavy-quark implementation and the Herwig 7 \texttt{MEPP2QQ} process. The simulated results agree well with the experimental data.
The $\rm D^0$-mesons used to tag the jets have a minimum transverse momentum $p_{T,D}>2$~GeV/$c$ for $5<p_{T,jet}^{ch}<15$~GeV/$c$ and $p_{T,D}>6$~GeV/$c$\ for $15<p_{T,jet}^{ch}<30$~GeV/$c$.
These kinematic cuts allow one to fully access the $z_{||}^{ch}$ distribution in $0.4<z_{||}^{ch}<1.0$ for both jet momentum intervals. In the lower $p_{T,jet}^{ch}$ interval, a pronounced peak at $z_{||}^{ch}\approx1$ is observed. This peak is populated by jets in which the $\rm D^0$-meson is the only constituent. Whereas, in case of the higher $p_{T,jet}^{ch}$ interval single-constituent jets are much rarer and the peak at $z_{||}^{ch}\approx1$ disappears. In general, as $p_{T,jet}^{ch}$ increases, the fragmentation becomes softer, a feature that has been observed also for inclusive jet measurements by ATLAS Collaboration~\cite{ATLAS:2011a}.

\begin{figure}[!h]
	\centering
	\begin{minipage}[b]{0.4\textwidth}
		\includegraphics[width=\textwidth]{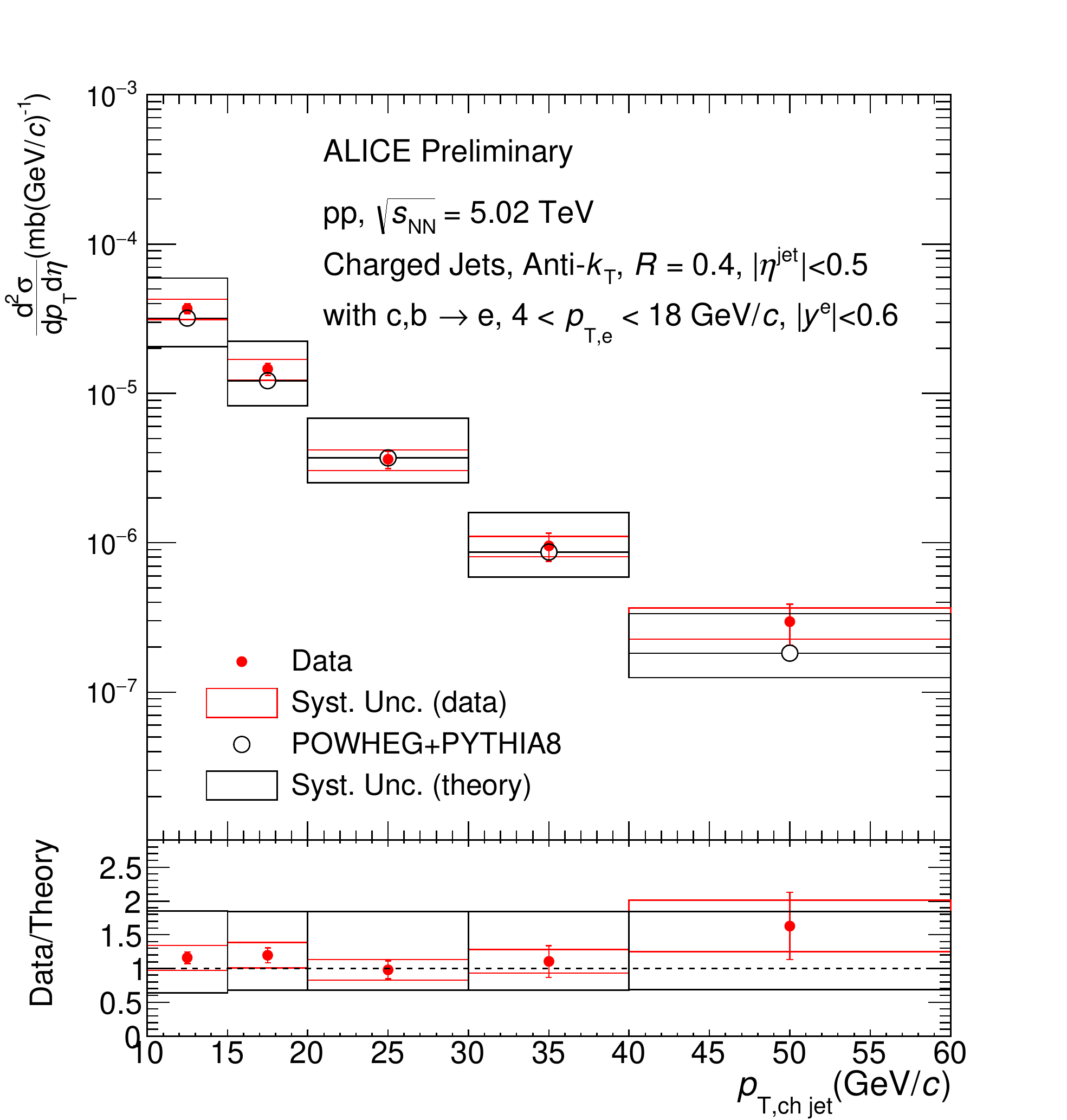}
		\caption{ The $p_T$-differential cross section of heavy flavor jet (tagged with HFe) in pp collisions at $\sqrt s = 5.02$ TeV, compared with the results of POWHEG+PYTHIA8.}
		\label{HFejet}
	\end{minipage}
\end{figure}

The Fig.~\ref{HFejet} depicts the $p_T$-differential cross section of heavy flavor jet in pp collisions at $\sqrt s = 5.02$ TeV. The identification is performed by HFe where the momentum range of HFe is $4<p_{T,e}<18$~GeV/$c$. The data are compared with POWHEG calculations and the comparison shows the best agreement between data and POWHEG calculations within their uncertainties.

\begin{figure}[!h]
	\centering
	\begin{minipage}[b]{0.44\textwidth}
		\includegraphics[width=\textwidth]{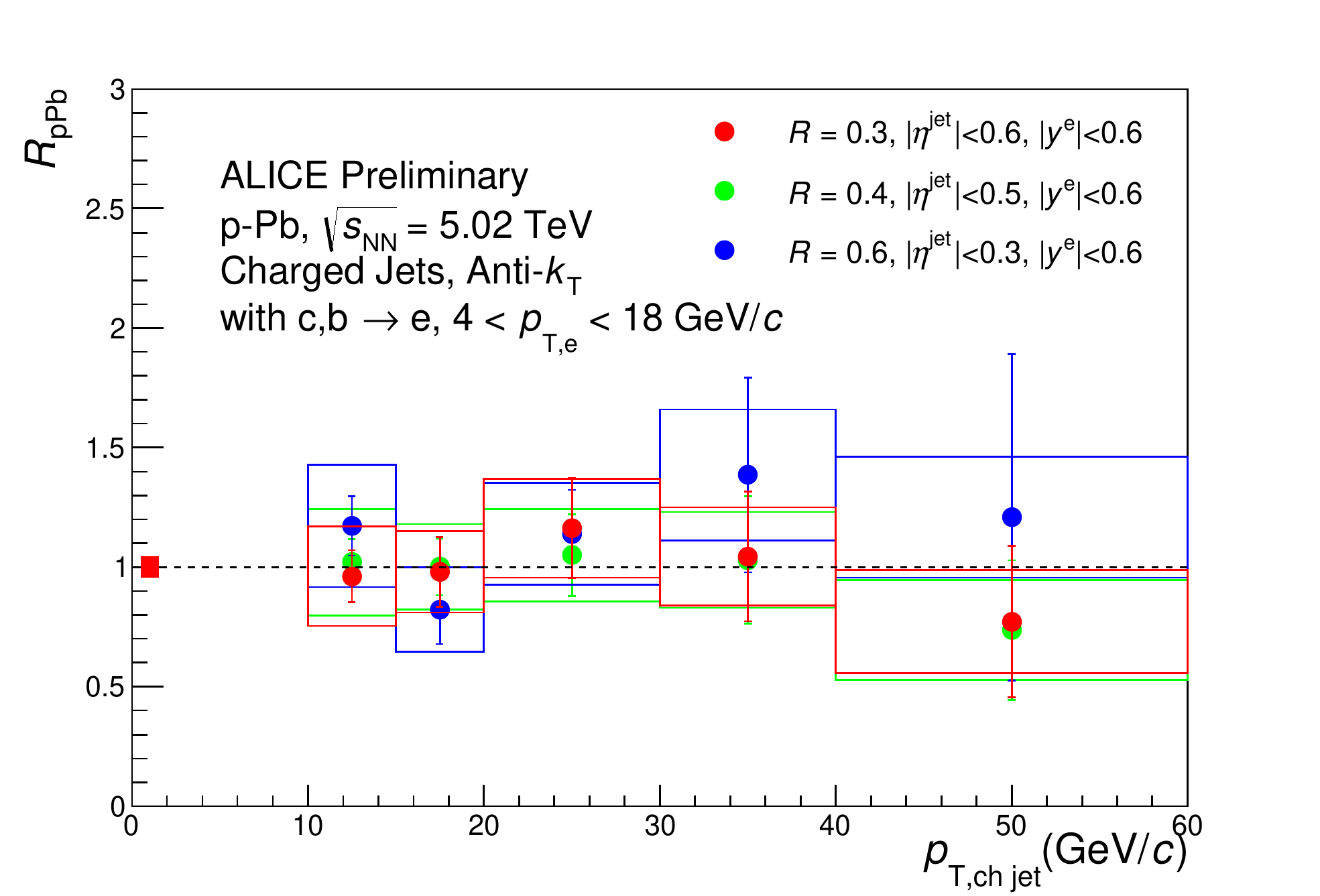}
		\caption{ Heavy flavor jet (tagged with HFe) nuclear modification factor ($R_{pPb}$) in p--Pb collisions at $\sqrt {s_{\rm NN}} = 5.02$ TeV.}
		\label{RpAHFejet}
	\end{minipage}
\end{figure}

\begin{figure}[!h]
	\centering
	\begin{minipage}[b]{0.36\textwidth}
		\includegraphics[width=\textwidth]{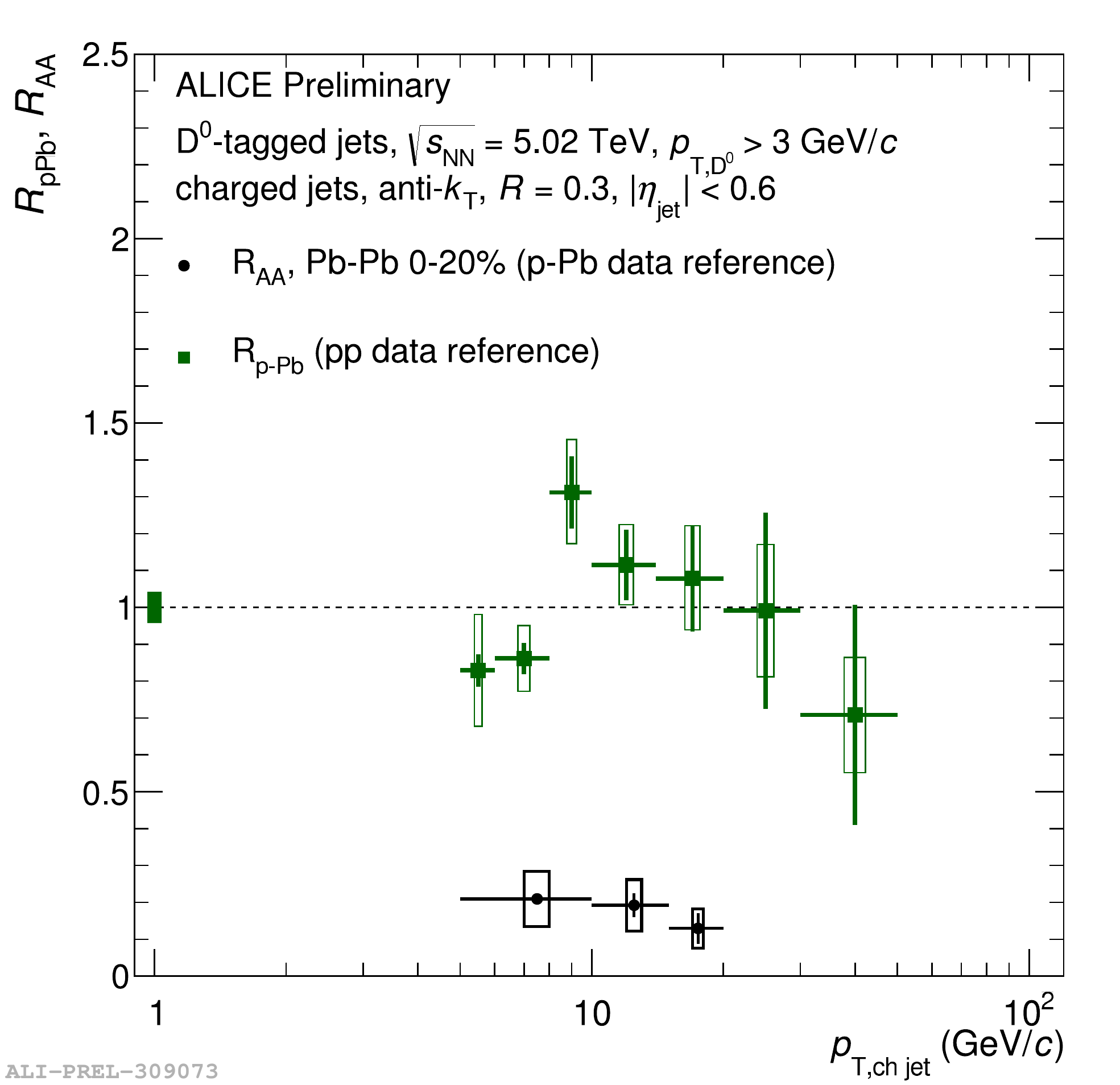}
		\caption{ The nuclear modification factor ($R_{pPb}$,$R_{AA}$) of charm jet (tagged with $\rm D^0$-meson) in p--Pb and Pb--Pb collisions at $\sqrt {s_{\rm NN}} = 5.02$ TeV.}
		\label{RAADjet}
	\end{minipage}
\end{figure}

The nuclear  modification  factor $R_{pPb}$ of HFe tagged jets as a function of $p_T$ in p--Pb collisions at $\sqrt {s_{\rm NN}} = 5.02$ TeV is shown in Fig.~\ref{RpAHFejet}. The $R_{pPb}$ is consistent with unity within the uncertainties over the entire $p_T$ range of the measurements. The production of heavy flavor jet is thus consistent with binary collision scaling of the reference spectrum for pp collisions at the same centre-of-mass energy. The measurements immediately suggest that there is a negligible initial state effect (CNM effect) present in case of heavy flavor jets. The suppression of the heavy flavor jet yield in Pb--Pb collisions is thus the final state effect induced by the produced hot medium as shown in the Fig.~\ref{RAADjet}.

The nuclear modification factor in Pb--Pb collisions is defined as:

\begin{eqnarray}
R_{AA} = \frac{1}{A}\frac{\rm d\sigma_{AA}/dp_T}{\rm d\sigma_{pp}/dp_T} 
\end{eqnarray}
where, $\rm d\sigma_{AA}/dp_T$ and $\rm d\sigma_{pp}/dp_T$ are the $p_T$-differential  production cross sections of a given  particle species in A-A and pp collisions, respectively, and A is the number of nucleons in the nucleus. In Fig.~\ref{RAADjet}, we display the $R_{AA}$ of charm jet (tagged with $\rm D^0$-meson) in Pb--Pb collisions at $\sqrt {s_{\rm NN}} = 5.02$ TeV for the centrality range 0-20\%. We also show here the $R_{pPb}$ for $\rm D^0$-meson tagged jets as reference. The measurements show the strong suppression of the $\rm D^0$-meson tagged jets in Pb--Pb collisions which ensures the in-medium energy loss of the charm jets in the produced hot medium.

\section{Summary}
\label{sumcon}

We presented the measurements of $p_T$-differential production cross section of $b$-jet in p--Pb collisions at $\sqrt{s_{\rm NN}}=5.02$ TeV and that of $c$-jet in pp collisions at $\sqrt s = 5.02, 7$ and $13$ TeV as measured by the ALICE experiment at the LHC. The measurements have been compared with the NLO pQCD calculations (POWHEG+PYTHIA6) and we find a good agreement between them. Along with that we reported the $z_{||}^{ch}$-differential cross section of $\rm D^0$-meson tagged jets for the jet momentum range, $5<p_{T,jet}^{ch}<15$~GeV/$c$ and $15<p_{T,jet}^{ch}<30$~GeV/$c$. The measurements agree well with the results from POWHEG heavy-quark implementation and the Herwig 7 \texttt{MEPP2QQ} process. The $z_{||}^{ch}$-differential measurement suggests softer heavy flavor jet fragmentation as $p_{T,jet}^{ch}$ increases, similar to inclusive jet~\cite{ATLAS:2011a} measurement. The measured cross section for heavy-flavor jet production in pp collisions at $\sqrt s = 5.02$ TeV agrees with POWHEG calculations within uncertainties.
Besides that, we also presented the measurements of nuclear modification factor of heavy flavor jets in p--Pb and Pb--Pb collisions  at $\sqrt {s_{\rm NN}} = 5.02$ TeV. The $R_{pPb}$ is consistent with unity implies insignificant initial state effect (CNM effect) for heavy flavor jets. The measured $R_{AA}$ of charm jet in Pb--Pb collisions is less than unity, showing a strong suppression of charm jet in the produced hot medium. The suppression is caused due to the energy loss of charm quark jets in the medium.

\end{document}